\documentclass{kluwer}    

\newdisplay{guess}{Conjecture}
\usepackage{graphicx}
\begin{document}                                                               
                    
\begin{article}
\begin{opening}         
\title{Constraining the Parameters of AGN Jets}
\subtitle{Comparisons with Herbig-Haro Jets}
\author{Silvano \surname{Massaglia}}  
\runningauthor{Silvano Massaglia}
\runningtitle{Parameters of AGN Jets}
\institute{Dipartimento di Fisica Generale dell'Universit\`a, 
Via Pietro Giuria 1, 10125 Torino, Italy}
\date{November 19, 2002}

\begin{abstract}
Comparing the properties AGN and Herbig-Haro jets can be a useful exercise for
understanding the physical mechanisms at work in collimated outflows
that propagate in such different environments. In the case of Herbig-Haro jets,
the presence of emission lines in the spectra and the continuous evolution of the
observation techniques greatly favor our knowledge of the physical parameters
of the jets instead, for AGN jets, the process of constraining the jet parameters is
hampered by the nature of the emission from these objects that is non-thermal.

I will discuss how one cannot directly constrain the basic parameters of extragalactic jets by
observations but  must treat and interpret the data either by statistical means or by comparing
observed and simulated morphologies in order to gain some indications on the values
of these parameters.
\end{abstract}

\end{opening}

\section{Introduction}

The main difficulty one faces when trying to comprehend the nature of
jets from Active Galactic Nuclei is due to the {\it absence of lines} in
the radiation spectrum of these objects.
In fact, their emission is typically non-thermal emission (synchrotron or inverse
Compton).
This very simple fact constitutes a sort of  ``original sin"
of AGN jets and is basically the reason why after almost five
decades since the discovery of the radio emission from Cygnus A (Jennison \&
Das Gupta 1953), observers and theorists are still debating about very basic 
questions such as
 extragalactic jets composition, i.e.
whether they are made of ordinary matter or
of electron-positron pairs.

Models of AGN jets, that consider the jet overall dynamics,
depend on {\it a minimum} of three parameters: the Lorentz factor, the Mach 
number,
the jet-to-ambient density contrast. Another crucial parameter
for jet modeling would be the
magnetic field  intensity, however one may assume, at a zeroth order
approximation, that the bulk kinetic energy density in the jet dominates on the
magnetic one and that the general behavior of the jet propagation can be
captured by a hydrodynamic model. In any event, none of the above
jet parameters can be {\it directly} constrained by observations.

In contrast, Herbig-Haro jets show copious emission
lines in their spectra, in the optical and infrared
bands mainly. Spectral lines give information on local temperature
and density, on the bulk velocity of the jet emitting matter and on the
presence of shocks along the jet. Moreover, the augmented angular resolution
and sensitivity achievable with the recent
telescopes (see Bacciotti, this issue) allow to investigate the
velocity structure across the jet and reveal the jet
characteristics close to the central source.

In this review I will present efforts to constrain the basic parameters in
extragalactic jets, both by statistical and numerical means, discuss the
role of these efforts in interpreting the FR I/FR II radio source dichotomy,
and made comparisons with observations and interpretations of Herbig-Haro
jets.  For an extended review on jet modelling, I remind the reader to Ferrari
(1998).

In the next Section I briefly introduce the bases of the Fanaroff-Riley classification
of extragalactic radio sources and recall the interpretations proposed;
in Section 3 I examine a method for deriving jet velocities; in Sections 4 and 5
I discuss the problem of constraining the jet Mach number , density and composition
and in Section 6  make comparisons between Herbig-Haro and AGN jets.

\section{The Fanaroff-Riley classification: Taxonomy}

\begin{figure}
\resizebox{\hsize}{!}{\includegraphics[angle=90]{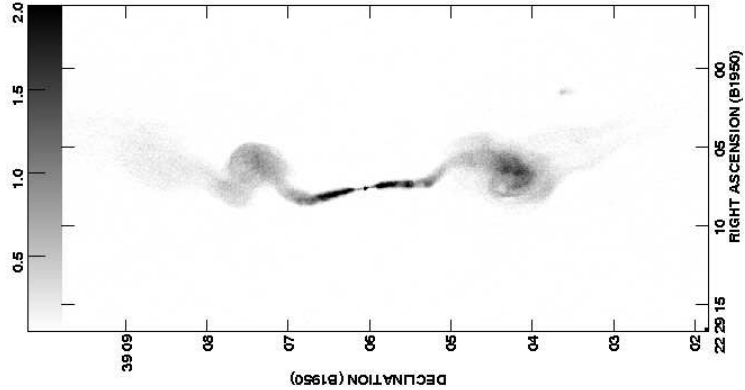}}
\vskip -1.5cm
\caption{VLA image of the FRI source 3C449 (Feretti et al. 1999)}

\label{fig:FRI}
\end{figure}

\begin{figure}
\resizebox{\hsize}{!}{\includegraphics{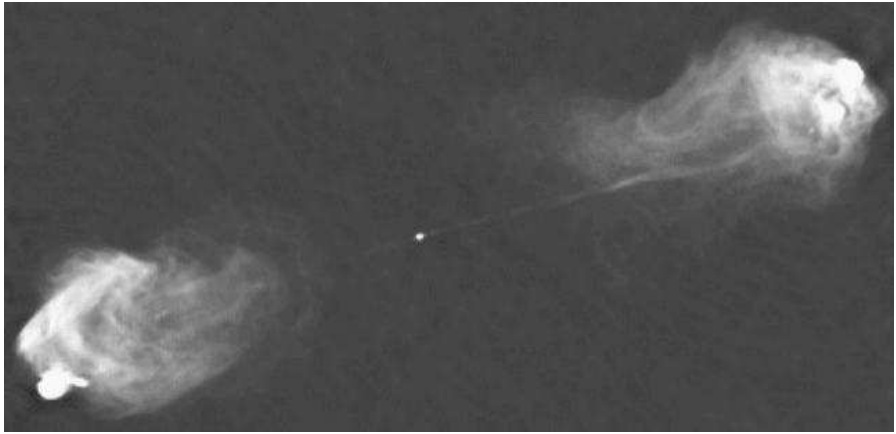}}
\caption{VLA image of the FRII source Cygnus A}

\label{fig:FRII}
\end{figure}

Historically, the extragalactic radio sources have been classified into two
categories (Fanaroff \& Riley 1974) based upon their radio morphology: a
first class of objects, preferentially found in rich clusters and hosted by
weak-lined galaxies,
 shows jet-dominated emission and two-sided jets and  was named FR I
(Fig. \ref{fig:FRI}); a second one, found isolated or in poor groups and
hosted by strong emission-line galaxies,
presents lobe-dominated emission and one-sided jets and was called FR II (or
``classical doubles") (Fig. \ref{fig:FRII}).
Besides morphology, FR I and FR II radio sources were discriminated in power
as well: objects below
$\sim 2 \times 10^{25} h_ {100}^2$ W Hz$^{-1}$ str$^{-1}$  were typically
referred as FR I sources. A perhaps
more illuminating criterion has been found by Owen \& Ledlow (1994) who 
plotted the radio luminosity
against the optical absolute  magnitude of the host galaxy:
they found the bordering line of FR I to
FR II regions correlating as $L_R \propto L_ {opt}^{1.7} $, i.e. in a luminous 
galaxy more radio power
is required to form a FR II radio sources.  This correlation is important 
since can be interpreted as an
indication that the environment may play  a crucial role in determining the 
source structure.
The above argument yields the basic question of the origin of FR I/FR II 
dichotomy, whether intrinsic
or ambient driven.

Two kind of explanations for the  FR I/FR II dichotomy can be found in the 
literature: {\it intrinsic} and
{\it extrinsic} interpretations (see the review by Wiita 2002). Among the {\it 
intrinsic} explanations,
the jet composition was invoked to interpret this dichotomy:  Celotti \& 
Fabian (1993)
argued that FR II radio jets were made of ordinary matter ($e^+-p$) and 
Reynolds et
al. (1996a) instead FR I jets (in  particular M87) were made of  $e^+-e^-$  
pairs. Another possibility
was considered by Wilson \& Colbert (1995) and Meier (1999) who suggested that 
FR II jets
originated from rapidly rotating black-holes; the structure of the accretion 
was also considered to explain
the dichotomy: according to Reynolds et al. (1996b) when accretion is  
advection dominated
(ADAF)  FR I jets results, while standard accretion disks generate FR II jets.
The  {\it extrinsic} explanations assume, apart from the total power, FR I and 
FR II jets are basically similar
close to the nucleus and that differences in the environment are able to 
destabilize, possibly via onset of turbulence in the flow, and decelerate FR I
jets effectively, while FR II jets succeed to propagate, nearly unchanged, up 
to the working surface to
produce the hot-spots (Bicknell 1995, Komissarov 1990, Bowman et al. 1996).

A possible clue for discriminating among these two kind of interpretations 
(Gopal-Krishna \&
Wiita 2000) was the observations of six HYbrid MOrphology Radio Sources 
(HYMORS)
that show FR I morphology on one side of the core and FR II morphology
on the other one: this is a clear indication that the environment play a basic
role in determining the radio source appearance.

\section{The Jet Velocity}

As discussed before, none of the basic parameters (Mach number, jet-ambient
density ratio and jet Lorentz factor) required for analytical and
numerical modeling of jets can be directly
constrained by observations of radio sources. Observers must therefore 
interpret their data relying
upon  statistical analyzes and look at these data through an assumed basic 
model.
This procedure is typically employed for the interpretation of the jet 
one-sideness and superluminal
velocities that are observed in several radio sources
at milliarcsecond scales using VLBI techniques.
The basic model adopted can be synthesized as follows: responsible for the 
emission is a distribution
of relativistic electrons that are advected at relativistic speed by the jet, 
assuming  homogeneous and isotropic jets
the flux ratio of the approaching jet to the receding one (Doppler boosting) 
is:
\begin{equation}
    \frac{F_{a}}{F_ {r}} =
    \left( \frac{1+\beta_{j} \cos \theta}{1- \beta_{j} \cos \theta} 
\right)^{2+\alpha}     \;,
    \label{eq:boost}
\end{equation}
 Where $ F_{a}, \ F_ {r}$ are the fluxes of the approaching and receding jet 
respectively, $\beta_{j}$
 is the jet bulk velocity in units of $c$, $\theta$ is the angle between the 
jet axis and the line of sight and
 $F \propto \nu^{-\alpha}$, with   $ \alpha \sim 0.5$ typically. If the
emission region moves towards the
 observer at the jet speed with the angle  $\theta$, photons emitted at a 
later time must travel a distance
 to the observer that is shorter than for photons emitted earlier, this makes 
the apparent time interval
 of the emission smaller and thus the apparent speed $ \beta_ {app}$ possibly 
larger than one:
 \begin{equation}
    \beta_ {app} =
    \frac{\beta_{j} \sin \theta}{1- \beta_{j} \cos \theta}      \;.
    \label{eq:superl}
\end{equation}
The observation of the jet-counterjet flux ratio  $ F_{a} /  F_ {r}$  and of  
$ \beta_ {app}$ in a source
would yield the actual jet speed and inclination. Unfortunately this is not 
possible in most cases: typically
the counterjet is too faint to be visible and superluminal proper motions are 
detected in
a relatively few objects only.

Jets in AGNs are clearly highly non isotropic objects,
accordingly  their appearance crucially depends on the viewing
angle. This question is discussed in the  {\it unified model}  for AGNs
(Antonucci 1993, Urry \& Padovani 1995).  According to this model the
appearance of an AGN is
dictated by the jet axis orientation with respect to the line of sight: 
radio-loud quasars and FR II radio
galaxies are the same kind of objects seen with increasing angle 
(high-luminosity unified scheme),
while BL Lac objects and FR I radio sources also belong to the same class 
viewed at increasing angles
(low-luminosity unified scheme).  Therefore one may ask
the following question: when observing
  with the highest possible resolution, i.e. with VLBI techniques,
a sample of different kind on AGNs jets showing mixed FR I and FR II
 morphologies at kpc-scales,
 is their jet Lorentz  factor distribution mirroring the various large scale 
morphologies or not?

\begin{figure}
\centering
\vskip -5cm
\resizebox{\hsize}{!}{\includegraphics{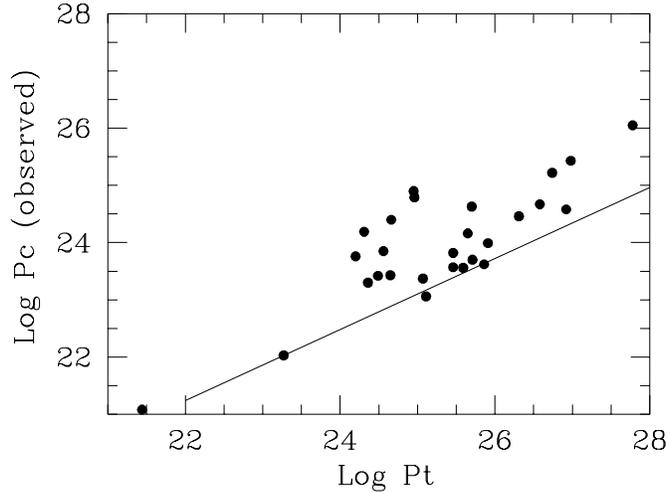}}
\caption{Observed arcsecond core radio power at 5 GHz against the total
radio power at 408 MHz. Solid line represents the correlation found by 
Giovannini
et al. (1988)}

\label{fig:giov1}
\end{figure}

\begin{figure}
\centering
 \vskip -5cm
\resizebox{\hsize}{!}{\includegraphics{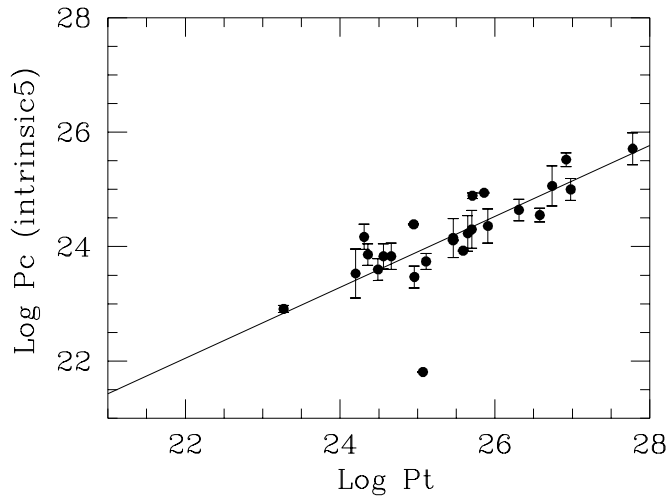}}
\caption{The solid line represents the correlation from Eq. (\ref{eq:corr}),
after the $P_{c}( 60^\circ)$ has been scaled.
Bullets show the intrinsic core power vs  the total radio power.
$\gamma=5$ has been assumed for deboosting
}
\label{fig:giov2}
\end{figure}

For attempting to answer  this question a possible way, as mentioned before, 
is to employ a statistical
analysis.  Giovannini et al. (1988) collected a sample of 187 radio galaxies, 
observed with the VLA, and
plotted their core (5 GHz) against total (408 MHz) power. Arguing that the 
objects
of this large sample had jets randomly oriented with the line of sight,  they
expected that  data were scattered about a
best-fit line that correlated core against total radio power and corresponded 
to the  mean
orientation angle of  $60^\circ$ to the line of sight. The correlation found 
was:
 \begin{equation}
    \log P_ {c}( 60^\circ)=0.62 \log P_{t} + 7.6     \;.
    \label{eq:corr}
\end{equation}
For investigating the jet properties close to the AGN,
Giovannini et al. (1994, 2001) examined a complete sample of 27 radio galaxies 
limited in flux observed with
 the VLBI. The sample included FR I and FR II sources in nearly the same 
quantity.
 They again plotted the observed core at 5 GHz,
 and affected by Doppler boosting, against the total radio power at 408 MHz, 
therefore not boosted, of
 the objects (Fig. \ref{fig:giov1}), with the correlation of Eq.  
(\ref{eq:corr})  given as a comparison.
Since both proper motions and jet-counterjet flux ratios were typically not 
available for the
sources of the sample, therefore Eqs.  (\ref{eq:boost}) and   
(\ref{eq:superl})  could not be solved
for $\theta$ and $\beta_{j}$, Giovannini et al. (2001)  fixed $\gamma$ and 
obtained $\theta$ from data,
once assuming $\alpha=0$, not a bad approximation for the core fluxes.
In this way they could constrain the Doppler factor $\delta \ (=[\gamma (1- 
\beta_{j} \cos \theta)]^{-1})$
and deboost the observed core power to obtain the intrinsic one, according to:
   \begin{equation}
     P_ {c \ obs}= P_{c \ intr} \times \delta^2     \;.
    \label{eq:deboo}
\end{equation}
The correlation  in Eq.  (\ref{eq:corr})  was scaled as well in the same way 
to obtain the intrinsic core
power and in Fig.  \ref{fig:giov2} one can see the result of this procedure 
after having set $\gamma =5$.
Repeating the same trick for different values of $\gamma$, Giovannini et al. 
(2001) found that the
correlation line went nicely through the data for $\gamma =3-10$, outside this 
interval the correlation
failed, as in  Fig.  \ref{fig:giov1}, {\it independently of the FR I/FR II 
type of the sources}.
Giovannini et al. (2001) concluded that FR I and FR II, despite showing 
different
kpc-scale morphologies, have all the Lorentz factors in the range 3-10 on the 
parsec scale.

\section{Jet Mach Number}

\begin{figure}
\centering
\resizebox{3in}{!}{\includegraphics{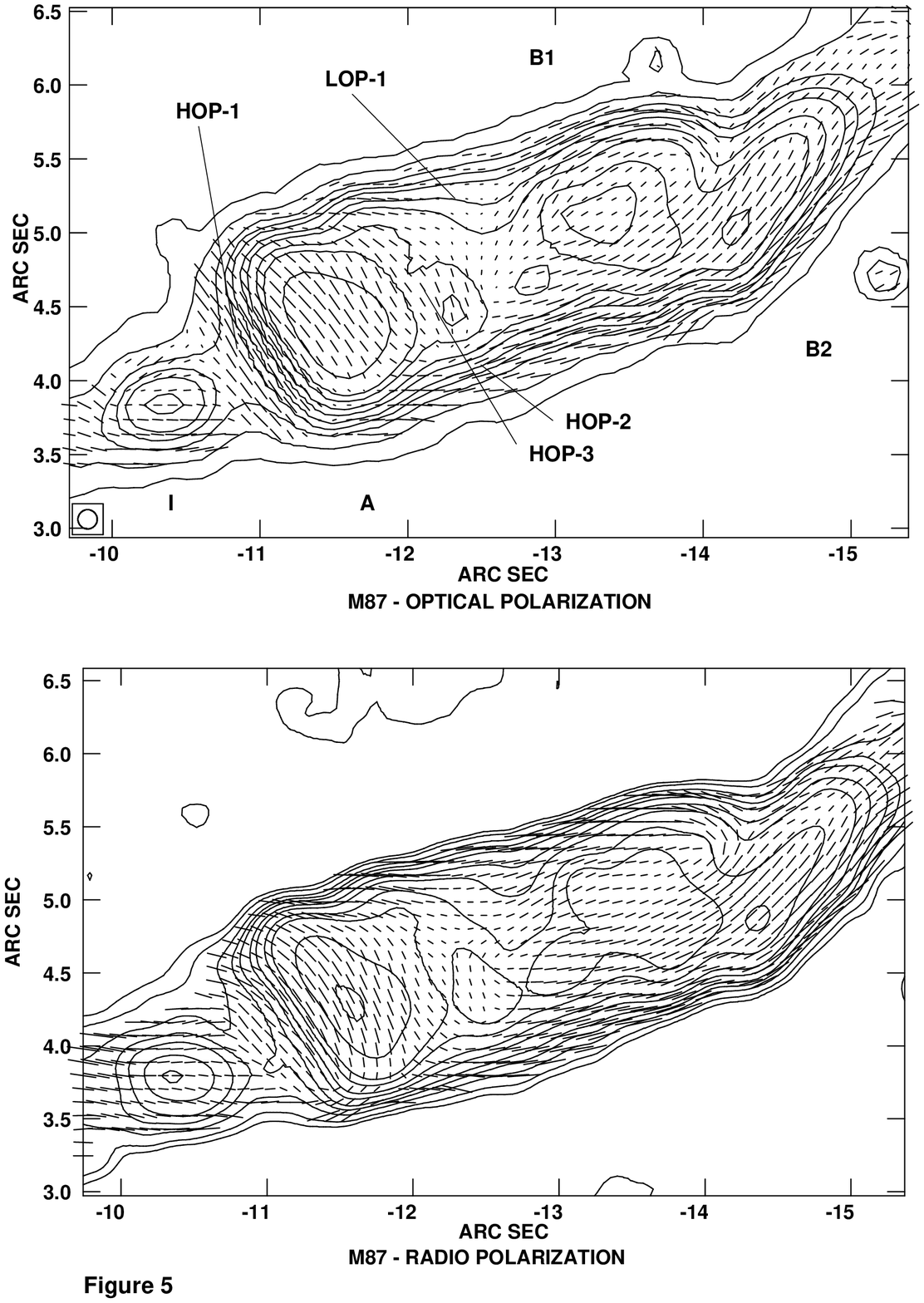}}
\vskip -0.9cm
\caption{Polarization maps of the M 87 jet (the projected magnetic field is 
given);
optical (top panel) and radio (bottom panel)}

\label{fig:pol}
\end{figure}

According to the above reasoning, both class of
AGN jets are relativistic, and thus supersonic,  on parsec scale. On
the kpc-scale
FR II jets may still be relativistic in most cases due to the observed 
one-sideness, if this
is interpreted as Doppler boosting. A clear indication of supersonic speed is 
the presence of
shocks in the FR I jets, that can be resolved in the transverse
direction, and the signature of shocks is the behavior of the polarization 
vector.  The M 87 jet
is among the best studied objects since originate from the closest AGN and 
detailed polarization maps
have been obtained in the radio and optical bands (Perlman et al. 1999). The 
magnetic
field is parallel to the jet axis and becomes perpendicular in the regions 
where we see
emission knots, such as knot A of the M 87 jet that is shown in detail in Fig. 
 (\ref{fig:pol}). This is
a clear indication of a shock compression of the field component along the 
shock extent.

\begin{figure}
\centering
\resizebox{2.5in}{!}{\includegraphics{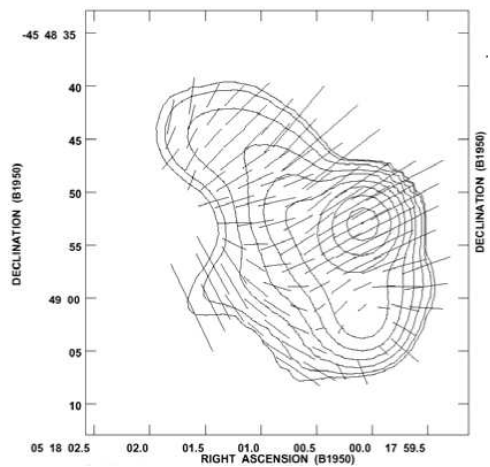}}
\caption{Polarization map of the western lobe of Pictor A (the electric field 
vector is given)
}

\label{fig:pol1}
\end{figure}

Jets of FR II radio sources are too faint to produce resolved polarization 
maps. Polarization
can be easily detected in the prominent radio lobes and hot spots, where the 
jet terminate into
the intergalactic medium. Again one expects a field direction  oriented along 
the lobe's border,
i.e. the working surface of a supersonic jet at its terminal point. In  Fig.  
(\ref{fig:pol1}) the
polarization map of the western hot-spot of Pictor A is shown (Perley et al. 
1997), and the projected magnetic field
is perpendicular to the electric field vector shown in the figure. The 
magnetic field results aligned
with the shocked working surface.

 \section{Jet Density and Composition}

We have seen that jets are likely relativistic on parsec scale, and that 
remain supersonic on
kpc scale, provided the direction and intensity of the magnetic field is 
correctly interpretated in terms of
shock compression.  A third parameter to constrain is the jet density, or more 
properly the jet-to-ambient density contrast.
For constraining this parameter one has to   rely upon  numerical simulations 
and compare qualitatively
the spatial {\it density}  distribution, resulting from the calculations,
with the observed  {\it brightness}  distributions.

\begin{figure}
\resizebox{\hsize}{!}{\includegraphics{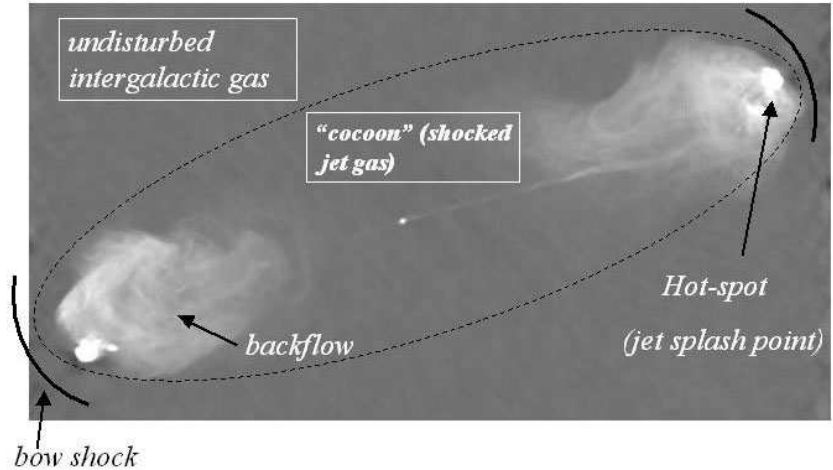}}
\caption{Observed brightness distribution in the FR II source Cygnus A
}

\label{fig:dia1}
\end{figure}

In  Fig.  \ref{fig:dia1}  I show the radio brightness
distribution of the FR II source Cygnus A and point out
the different regions of interest for the model interpretation: the radio
hot-spots are the regions where the jets terminate (``splash  points"), the
lobes are the manifestation of the back-flow, and the diffuse region between
the lobes is the  ``cocoon", typically invisible in radio but evident
in X-ray (see Smith et al. 2002).

\begin{figure}
\resizebox{\hsize}{!}{\includegraphics{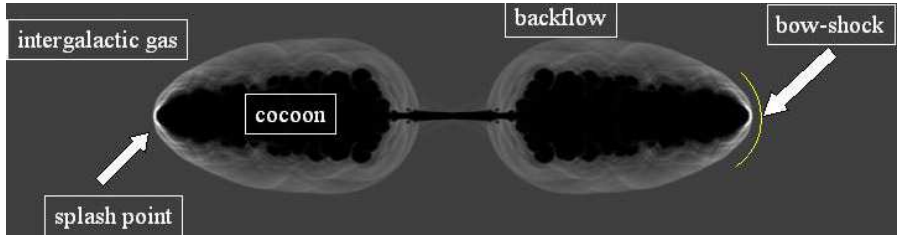}}
\caption{Simulated density distribution for a relativistic, supersonic and 
underdense
jet
}

\label{fig:dia2}
\end{figure}

When simulations of supersonic, underdense jets are carried out
(see e.g. Massaglia et al. 1996)
the overall picture that emerges from these studies can be summarized
as follows (see Fig.  \ref{fig:dia2}): the deceleration of the jet flow at its head is
accomplished through the formation of a strong shock (Mach disk), which
partially thermalizes the jet bulk kinetic energy; the overpressured
shocked jet material forms a back-flow along the sides of the jet and
inflates a cocoon whose size becomes more prominent as  the density
ratio between jet and ambient material decreases; finally, a second
shock (bow-shock) is driven into the external medium. Simulations of
supersonic, overdense jets yield qualitatively different morphologies that
are close to Herbig-Haro jets. Therefore AGN jets, at least those found in
FR II sources, are underdense with respect to external medium and
the density ratio can be very small  ($\sim 10^{-5}$) in case pair plasma
jets.

The question of the jet density reminds the problem of establishing whether
AGN jets are made of proton and relativistic electrons or instead of relativistic
electron-positron pairs.  This of the jet composition is a long outstanding
and yet unsolved problem for the simple reason that
electrons and positron emit synchrotron radiation of identical spectrum and
linear polarization.  A possible observational clue for discriminating between
$e^- -p$ and $e^+ - e^-$ jets is the detection of radio circular polarization (CP).
In fact (see Wardle et al. 1998),  the large linear polarization degree observed
in AGN jets at parsec scales has survived the internal Faraday rotation process,
that is mainly due to electrons with $\gamma_ {\rm min} < 100$. Therefore    $e^- -p$
jets should have relativistic electrons with     $\gamma_ {\rm min} \ge 100$,
while   $e^+ - e^-$ jets do not suffer Faraday rotation since electron and
positron gyrate in opposite directions and  thus  $\gamma_ {\rm min} \ge 1$.
Values of  $\gamma_ {\rm min} \ll 100$ should be a clear signature of  $e^+ - e^-$ jets.
Low-energy electrons can yield to {\it Faraday conversion} of linear into circular polarization,
besides producing  Faraday rotation.  For an  $e^+ - e^-$ jet internal Faraday rotation
should not be present, as mentioned before, and a small but
appreciable fraction (below 1\%) of the linear
polarization should be converted into CP by low energy electrons and positrons since
this mechanism is sign-independent, being proportional to the electron gyrofrequency
to the square.  Observation of CP should be thus be a signature of the presence of
low-energy emitting particles that do not cause Faraday rotation and  maintain
a high degree of LP, therefore should be $e^+ - e^-$ plasma.

CP has been indeed observed   in about 20 AGN jets (Wardle \& Homan 2001) and we should have
a clear indication on the  $e^+ - e^-$ nature of the jet composition. Unfortunately, there are
other possible origins of CP, besides Faraday conversion: intrinsic CP, scintillation and coherent
radiation mechanisms. Therefore the question of the jet composition remains open.

Authors have attempted different ways for solving this important and basic problem.
Scheck et al. (2002) have simulated relativistic $e^+ - e^-$ and  $e^- -p$ jets of given
kinetic luminosities and jet-ambient density ratios: they have found that
{\it both the morphology
and the dynamical behavior is almost independent of the assumed jet composition}.

\section{AGN and HH Jets}
AGN and Herbig-Haro jets (see a recent review by Reipurth \& Bally 2001) share
some similarities but many basic differences,
besides the different scale powers and sizes. Both are collimated outflows that
originated from accretion processes and are accelerated, possibly by MHD mechanisms.
Since jet velocity must be of the order of the escape velocity from the central
gravitational well, AGN jets are likely to be relativistic, at least in their initial part,
while HH jets have velocities comparable with stellar winds.
Both kind of jets appear knotty and terminate into the ambient medium forming
structures whose morphology that can be interpreted as bow-shocks, that indicates
supersonic bulk velocity in the jets.

HH jets are observed in the optical and infrared bands where they emit
optically thin thermal radiation.
Their spectra are rich of emission lines, mainly of hydrogen, sulfur and oxygen with
intensities, ratios and profiles are clearly indicative of shocks, since these observational
values can be
related to the predictions from plane-parallel and bow-shock models. Combining the data
of shock velocities, Doppler shifts of the emission lines and proper motions it is
possible to constrain the jet velocity, density and temperature (i.e. the Mach number).
Typically, the jet velocity is $\sim 200 - 400$ km$^{-1}$, density  $\sim 10^3 -10^4$ cm$^{-3}$
and Mach number $\approx 20-40$.

AGN jets instead emit continuous, non-thermal radiation in a wide frequency range that
goes from the radio up to the X-ray bands; these jets are likely relativistic  on parsec scale
but may decelerate down to non-relativistic speeds at kpc scales, at least in the case of FR I
objects, but are likely to remain supersonic both FR I and FR II jets.  The basic reason why
shocks lead to thermal emission in HH objects and non-thermal radiation in AGN jets is
due to about 6-8 orders of magnitude ratio in the ambient density: in the case of HH
jets propagate in the high density ambient of molecular clouds,
thus shocks can heat the matter and produce radiative losses, while extragalactic jets
propagate in the tenuous intergalactic or intracluster medium therefore shocks,
that can be considered with good approximation adiabatic, have the main effect to
accelerate relativistic particles via Fermi-like processes of the first kind
that, in turn, yield synchrotron radiation in the shocked ambient magnetic field and the
only possible signature we observe is the reorientation of the polarization vector of the
radiation.

\end{article}
\end{document}